%% file: GWB-KinAnisPTAs_V3.0.tex
\documentclass[a4paper,11pt]{article}
\usepackage{aaskaiid}
\def\hb{{\boldsymbol{H}}}
\usepackage{url}
\usepackage{soul}
\usepackage{orcidlink}
\def\Tr{{\rm Tr}}

\def\be{\begin{equation}}
\def\ee{\end{equation}}

\title{Probing Kinematic Anisotropies in the Stochastic Gravitational Wave Background with the SKA
 }
\ShortTitle{SGWB kinematic anisotropies}

\author[1]{N.~M.~J.~Cruz \orcidlink{0009-0003-4719-2126}}
\ShortName{N.~M.~J.~Cruz et al.} 
\author[1]{Ameek Malhotra \orcidlink{0000-0001-8346-9995}}
\author[1,2]{Gianmassimo Tasinato \orcidlink{0000-0002-9835-4864} }
\author[1]{Ivonne Zavala \orcidlink{0000-0002-5589-9928}}

\affiliation[1]{ 
Physics Department, Swansea University, SA2 8PP, UK
}
\emailAdd{nmjc1209.at.gmail.com}
\affiliation[2]{Dipartimento di Fisica e Astronomi\textbf{}a, Universit\`a di Bologna, INFN, Sezione di Bologna, viale B. Pichat 6/2, 40127 Bologna, Italy
}
\emailAdd{ameek.malhotra.at.swansea.ac.uk}
\emailAdd{g.tasinato2208.at.gmail.com}
\emailAdd{e.i.zavalacarrasco.at.swansea.ac.uk}

\abstract{
Pulsar Timing Arrays (PTAs) and astrometric surveys provide complementary probes of the nanohertz stochastic gravitational-wave background (SGWB). A primary target is the kinematic dipole induced by the Solar System's motion, whose detection would confirm the SGWB's cosmological origin. We forecast the sensitivity of the Square Kilometre Array Observatory (SKAO) using optimal estimators and Fisher techniques, considering both the baseline AA4 configuration and an optimistic $\sim 1000$-pulsar scenario, including the potential gain from combining with \emph{Gaia}-like astrometry. While SKAO will substantially improve constraints on SGWB anisotropies, even joint analyses remain below the sensitivity required to detect the kinematic dipole, motivating new observational and analysis strategies to fully exploit this signal.

}

\begin{document}

\include{journal-names}

\maketitle

\section{Introduction and Motivation}

Pulsar Timing Arrays (PTAs) provide a unique window onto the nanohertz gravitational-wave (GW) spectrum. By precisely timing an ensemble of millisecond pulsars, PTAs can detect the spatially correlated perturbations in pulse arrival times induced by passing GWs \citep{Hellings:1983}. Recent PTA observations have revealed evidence for a common-spectrum stochastic process \citep{NANOGrav:2023gor,Reardon:2023gzh,Xu:2023wog,EPTA:2023fyk,InternationalPulsarTimingArray:2023mzf}, marking the onset of GW astronomy in the nanohertz band.

The  Square Kilometre Array Observatory (SKAO) will greatly enhance PTA capabilities.  By observing hundreds of millisecond pulsars at sub-100~ns timing precision and with nearly all-sky coverage, the SKAO will improve GW sensitivity by   an order of magnitude. This will enable  a detailed characterization of the stochastic GW background, and also stringent tests of fundamental physics.

Complementary to PTAs, astrometric surveys such as \emph{Gaia} \citep{Gaia:2016zol,Gaia2023}, \emph{Theia} \citep{Theia:2017xtk}, and the \emph{Nancy Grace Roman Space Telescope} \citep{sanderson2019astrometrywidefieldinfraredspace} provide an independent and powerful probe of nanohertz GWs \citep{Perna01.2026.SKA}. Gravitational waves propagating between the Earth and distant sources induce correlated, time-dependent deflections in the apparent positions of those sources. Existing data from \emph{Gaia} and VLBI already constrain the stochastic GW background at frequencies $f \lesssim 10^{-9}$ Hz, while future missions could extend sensitivity to the $f \sim 10^{-4}$–$10^{-7}$ Hz range \citep{Jenkins01.2026.SKA, Capelo01.2026.SKA}, thereby complementing PTA observations \citep{Gwinn:1996gv,Titov_2011,Darling:2018hmc,Takahashi01.2026.SKA}.
On comparable scales, ultra-low-frequency GWs can affect the apparent shapes of galaxies, producing the so-called \emph{cosmic shimmering} effect \citep{Mentasti:2024fgt} and influencing intrinsic galaxy alignments (see \cite{Chisari:2025gsy} for a comprehensive review). Such effects could be detected by current and forthcoming surveys such as \emph{Euclid}\footnote{\url{https://www.esa.int/Science_Exploration/Space_Science/Euclid}.}.
Hence, the synergy between PTAs, astrometric measurements, and large-scale structure observations will open new avenues for probing the low-frequency GW Universe. Maximizing the scientific return from these complementary approaches will be crucial in the coming decade.

A stochastic gravitational-wave background (SGWB) may originate from astrophysical sources, e.g., supermassive black hole binaries \citep{Sesana:2008mz,Burke-Spolaor:2018bvk}, or from cosmological processes such as inflation or phase transitions \citep{Caprini:2018mtu, Pasechnik01.2026.SKA}. While astrophysical backgrounds are expected to exhibit strong intrinsic anisotropies due to source clustering \citep{Cornish:2013aba,Taylor:2020zpk}, (see \cite{Pol:2022sjn,NANOGrav:2023tcn,Depta:2024ykq} for recent studies on their detectability), cosmological signals are intrinsically nearly isotropic, with anisotropies of order $\sim 10^{-5}$ (see~\cite{Bartolo:2019oiq,Contaldi:2016koz}).

The most important source
of anisotropy for a cosmological SGWB  is
 a {\it kinematic dipole} induced by the motion of the Solar System with respect to the GW rest frame. {This is characterized by the dimensionless parameter $\beta = v/c$ that quantifies the velocity of our motion with respect to the cosmological rest frame. Its magnitude and direction are accurately inferred from the CMB temperature dipole and correspond to $\beta \simeq 1.23 \times 10^{-3}$ towards $(l,b) = (264^\circ,48^\circ)$ in galactic co-ordinates~\citep{Smoot:1977bs,Kogut:1993ag,Planck:2013kqc}. If the SGWB is cosmological and statistically isotropic in its own rest frame, the observer’s motion induces a dipolar\footnote{{Higher multipoles are also generated, at order $
 \beta^2$ and beyond. Since the value of $\beta$ inferred from the CMB is quite small, these higher order effects are highly suppressed.}} modulation of the GW energy density through Doppler and aberration effects, with amplitude of order $\mathcal{O}(\beta)$ \citep{Cusin:2022cbb,LISACosmologyWorkingGroup:2022kbp,Tasinato:2023zcg}. In this case, the kinematic dipole of the SGWB is fully characterized by the same parameter $\beta$ measured in the CMB. Any intrinsic anisotropies of the SGWB would instead have
 a different angular and frequency dependence, and are conceptually distinct from the kinematic contribution}. Currently, forecasts
exist to detect such dipole in the 
radio band with SKA \citep{Bengaly:2018ykb}, but a measurement
with GW would be instrumental
as {\it independent probe} of kinematic
anisotropies -- possibly also clarifying some 
existing tensions between CMB measurements
and large-scale structure surveys (see e.g. 
\cite{Yoo:2025qdq} for a recent account).

Hence the 
detection of the kinematic dipole is a primary target for next-generation PTAs and SKAO-era astrometry, offering both a consistency check for cosmological SGWB models and a measurement of the observer's velocity relative to the GW rest frame.

This subject is the scope of this
chapter, which aims to summarize our works \cite{Tasinato:2023zcg,Cruz:2024svc,Cruz:2024esk,Cruz:2024diu}
and to extend the corresponding
forecasts to the SKAO survey. 
The chapter is organized as follows. In Section~\ref{sec_theory}, we review the theoretical motivations and present the formalism necessary to describe kinematic anisotropies in the SGWB. Section~\ref{sec_fisher} applies this framework to forecast the sensitivity of SKAO observations, both alone and in combination with astrometric data, to detect these anisotropies. 
In Section~\ref{sec_conclusions}
we summarize our findings, discussing their implications and possible ways forward.

\section{Kinematic anisotropies and
GW experiments in the nanohertz band}
\label{sec_theory}

In this section we collect the theoretical expressions relevant for our forecasts.  
Let $I(f,\hat n)$ denote the specific intensity of the stochastic gravitational-wave background (SGWB) at frequency $f$ in direction $\hat n$.
Under a Lorentz boost with velocity ${\bf v}=c\,\beta\,\hat v$, and to first order in $\beta$, the frequency and direction transform as \citep{Cusin:2022cbb}:
\begin{align}
f' &= f \, (1 + \beta\, \hat n\!\cdot\!\hat v), \\
\hat n' &= \hat n + \beta \big[(\hat n\!\cdot\!\hat v)\hat n - \hat v\big].
\end{align}
The intensity transforms according to
\begin{equation}
I'(f',\hat n') = \frac{f'^3}{f^3}\, I(f,\hat n).
\end{equation}
To first order in $\beta$, the fractional anisotropy is then
\begin{equation}\label{eq:Doppler}
\frac{\delta I(f,\hat n)}{\bar I(f)} = 
\beta (1 - n_I)\, \hat n\!\cdot\!\hat v,
\qquad
n_I \equiv \frac{d\ln \bar I}{d\ln f}.
\end{equation}

\subsection{Pulsar timing anisotropies}
\label{sec_theoryPTA}

PTAs measure the timing residuals induced by passing gravitational waves.  
For a pulsar in direction $\hat x$ and a GW propagating along $\hat k$, the timing residual induced by the metric perturbation $h_{ij}$ is given by (see e.g. \cite{Anholm:2008wy})
\begin{equation}\label{eq:timing}
r(t,\hat x) = \frac{1}{2}\,\frac{\hat x^i \hat x^j}{1 + \hat k\!\cdot\!\hat x} 
\left[ h_{ij}(t_{\rm e}, 0) - h_{ij}(t_{\rm p}, x) \right],
\end{equation}
where $t_{\rm e}$ and $t_{\rm p}$ denote, respectively, the times of reception and emission.  
The \emph{Earth term} (first term inside the bracket) is coherent across all pulsars, producing correlated timing residuals that allow the SGWB to be detected statistically.

Assuming a stochastic background with power spectrum $S_h(f)$, the cross-correlation between the residuals of pulsars $a$ and $b$, separated by an angle $\cos \theta_{ab} = \hat x_a \cdot \hat x_b$, reads
\begin{equation}\label{eq:xcorr}
\langle r_a r_b \rangle = \frac{3 H_0^2}{32\pi^2 f^3}\, \Omega_{\rm GW}(f)\, \Gamma_{ab}(f),
\end{equation}
where $\Gamma_{ab}(f)$ is the \emph{overlap reduction function} (ORF).  
For an isotropic background with tensor polarizations, $\Gamma_{ab}$ reduces to the Hellings--Downs curve:
\begin{equation}
\Gamma_{ab}^{(0)}(\theta_{ab}) = \frac{3}{2}\,y_{ab}\ln y_{ab} - \frac{y_{ab}}{4} + \frac{1}{2}, 
\qquad 
y_{ab} = \frac{1 - \hat x_a \cdot \hat x_b}{2}.
\end{equation}

\begin{figure}
    \centering
    \includegraphics[width=0.6\linewidth]{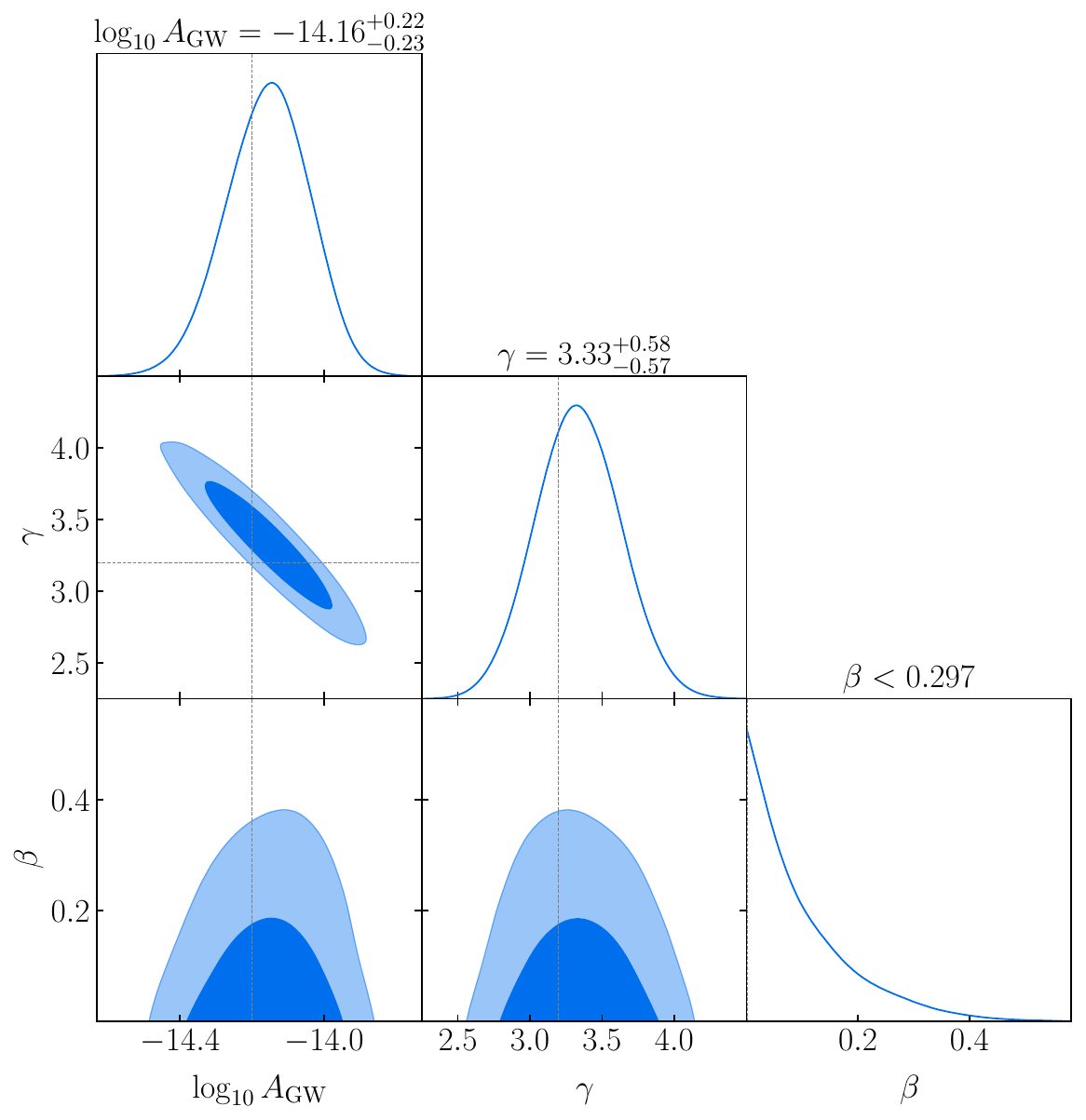}
    \caption{Parameter distributions and $95\%$ credible intervals for the SGWB amplitude, spectral index $\gamma$ and dipole magnitude~$\beta$ (from~\cite{Cruz:2024svc}). The gray dashed lines denote the median values of the amplitude and tilt obtained from the {NANOGrav} 15-year isotropic analysis~\cite{NANOGrav:2023gor}.}
    \label{fig:dipoleNg15}
\end{figure}

A moving observer detects a dipole-modulated SGWB according to Eq.~\eqref{eq:xcorr}.  
The corresponding ORF can be expanded as
\begin{equation}
\label{eq:Orf_upto_dipole}
\Gamma_{ab}(f) = \Gamma_{ab}^{(0)} + \beta (n_I -1)\, 
\Gamma_{ab}^{(1)}(\hat v) + \mathcal{O}(\beta^2),
\end{equation}
where $\Gamma_{ab}^{(1)}$ encodes the directional dependence of the correlation pattern and $n_I$ defined in Eq.~\eqref{eq:Doppler}.
%
Explicitly \citep{Cusin:2022cbb,Tasinato:2023zcg,Cruz:2024svc},
\begin{align}
\Gamma_{ab}^{(1)}(\hat v) &= 
\int \frac{d^2\hat k}{4\pi}\,
(\hat k\!\cdot\!\hat v)\,
F_a^{ij}(\hat k)\, F_{b,ij}(\hat k), \\
&= \left(\frac{1}{12} + \frac{y_{ab}}{2} + \frac{y_{ab}\ln y_{ab}}{2(1-y_{ab})}\right)
(\hat v\!\cdot\!\hat x_a + \hat v\!\cdot\!\hat x_b),
\end{align}
where the single-pulsar response function is
\begin{equation}
F_a^{ij}(\hat k) = 
\frac{1}{2}\,\frac{{\hat x_a^i \hat x_a^j }}{{1 + \hat k\!\cdot\!\hat x_a}}.
\end{equation}
The expected dipolar modulation of the timing-residual correlations  $C_{ab}$ scales as
\[
\frac{\Delta C_{ab}}{C_{ab}} \simeq 
\beta (n_I -1)\,
\frac{\Gamma_{ab}^{(1)}}{\Gamma_{ab}^{(0)}},
\]
which can reach ${\cal O}(10^{-3})$ for $\beta \simeq 10^{-3}$ and $n_I$ of order unity.  
Although this signal lies below the current PTA sensitivity, see Fig.~\ref{fig:dipoleNg15} for the upper limit on $\beta$ obtained from current data~\citep{Cruz:2024svc}, it may become detectable in the SKAO era, when hundreds of pulsars will be monitored with sub-100~ns timing accuracy.

\subsection{Astrometric anisotropies}
\label{sec_astronly}

Gravitational waves also induce correlated apparent proper motions in the sky positions of distant sources: see e.g. \cite{Linder_Astro,Braginsky:1989pv,Fakir_astro:1994,Pyne:1995iy,Kaiser:1996wk,Gwinn:1996gv,Kopeikin:1998ts,Jaffe:2004it,Schutz_2009,Book:2010pf,Titov_2011,Moore:2017ity,Klioner:2017asb,Darling:2018hmc,Mihaylov:2018uqm,Jaraba:2023djs}.  
Also these effects are sensitive to GWs in the nanohertz band, as in pulsar timing arrays.  
For a GW perturbation $h_{ij}$ propagating along $\hat k$, the induced astrometric deflection of a source in direction $\hat n$ is
\begin{equation}
\delta n_i(\hat n) = 
\mathcal{R}_{ikl}(\hat n,\hat k)\, h_{kl},
\qquad
\mathcal{R}_{ikl} = 
\frac{n_k}{2}\left[\frac{(n_i + k_i)n_l}{1+\hat n\!\cdot\!\hat k} - \delta_{il}\right].
\end{equation}

The two-point correlation of these deflections defines an astrometric overlap function $H_{ij}$, analogous to $\Gamma_{ab}$:
\begin{equation}
\langle \delta n^i(\hat n)\, \delta n^j(\hat m) \rangle 
\propto H_{ij}^{(0)}(\hat n,\hat m) 
+ \beta (1 - n_I)\, H_{ij}^{(1)}(\hat n,\hat m,\hat v),
\end{equation}
where $H_{ij}^{(0)}$ is the astrometric analog of the Hellings--Downs curve, sensitive to the isotropic SGWB, while $H_{ij}^{(1)}$ captures the kinematic dipole due to our motion relative to the GW rest frame.  
Introducing $y = \frac{1 - \hat n \cdot \hat q}{2}$, one finds \citep{Book:2010pf}:
\begin{align}
H^{(0)}_{ij}(\hat n, \hat q) = 
\frac{\pi}{3(1-y)^2}\left(1 - 8y + 7y^2 - 6y^2 \ln y \right)
\left[(2-2y)\delta_{ij} - n_i n_j - q_i q_j - q_i n_j + (1-2y) q_j n_i\right].
\end{align}

The dipolar correction can be written as
\begin{align}
H^{(1)}_{ij} = a_1 (A_i C_j + B_i A_j) + a_2 (B_i C_j - A_i A_j),
\end{align}
where
\begin{align}
\mathbf{A} &= \hat n \times \hat q, \quad
\mathbf{B} = \hat n \times \mathbf{A}, \quad
\mathbf{C} = -\hat q \times \mathbf{A},
\end{align}
and $a_1$, $a_2$ are scalar functions of the angles between $\hat n$, $\hat q$, and $\hat v$ (see \cite{Cruz:2024diu} for the full expressions).

\subsection{Joint PTA–astrometry correlations}

Joint PTA–astrometry analyses, particularly in the SKAO and \emph{Gaia} era, can exploit cross-correlations between $r_a$ and $\delta n_i$ (see e.g. the recent analysis~\cite{Vaglio:2025tex}), potentially enhancing the detectability of the kinematic dipole.  
The cross-correlation between a star’s deflection $\delta n_i(\hat n,t)$ and a pulsar’s timing residual $r_a(\hat x_a)$ reads
\begin{align}
\langle \delta n_i(\hat n)\, r_a(\hat x_a) \rangle 
= \frac{3 H_0^2}{64 \pi^3}
\int df\, \frac{\bar{\Omega}_{\rm GW}(f)}{f^3}
\left[ K_i^{(0)}(\hat n, \hat x_a) 
+ \beta (4 - n_\Omega)\, K_i^{(1)}(\hat n, \hat x_a, \hat v) \right],
\end{align}
where $K_i^{(0)}$ and $K_i^{(1)}$ trace the isotropic and kinematic components of the SGWB, respectively, {{and $ n_{\Omega} \,\equiv\,d \ln \Omega_{\rm GW}/d \ln f $ is the energy density spectral
tilt}}. Again, we refer
the reader to \cite{Cruz:2024diu} for full analytical expressions for these quantities. 
Building on these theoretical framework,
we now develop forecasts for understanding
 whether the kinematic dipole of the SGWB can be detected with SKAO. 

\section{Fisher forecasts}
\label{sec_fisher}

We now discuss applications and forecasts based on the formalism developed above. 
The monopole and dipole overlap reduction functions (ORFs), together with the cross-correlation functions, provide a unified and covariant framework for both PTA-only and joint astrometry–PTA analyses.  
These quantities form the key ingredients of Fisher-matrix forecasts where the expected parameter uncertainties are then obtained from the curvature of the log-likelihood around its maximum:
\begin{align}\label{deffim}
    {F}_{\alpha \beta} = \left\langle 
    -\frac{\partial^2 \ln \mathcal{L}}
    {\partial \Theta_\alpha\,\partial \Theta_\beta}
    \right\rangle,
\end{align}
Given a parameter vector $\hat{\Theta} = \{\Theta_\alpha\}$ describing the quantities of interest, the Fisher matrix sets a lower bound on the uncertainities with which these parameters can be measured \mbox{$\Delta \Theta_{\alpha,\rm min} = \sqrt{[F^{-1}]_{\alpha \alpha}}$}.



As in \cite{NANOGrav:2023gor}, we model 
the SGWB intensity profile as a power law:
\begin{equation}
I(f) = \frac{A_{\rm GW}^2}{2 f_{\rm ref}} \left( \frac{f}{f_{\rm ref}} \right)^{2-\gamma}, 
\end{equation}
with \footnote{{Using the conventions of \cite{Cruz:2024svc}, $\Omega_{\rm GW}(f) = \frac{2 \pi f^2}{3 H_0^{2}}A_{\rm GW}^2 \left( \frac{f}{f_{\rm ref}} \right)^{3-\gamma}$.}}$A_{\rm GW}$ controlling the amplitude, $\gamma$ the spectral slope, and $f_{\rm ref} = 1/$yr. We plan to use data to measure $A_{\rm GW}$, $\gamma$, 
as well as the kinematic dipole amplitude
$\beta$.

The analytical expressions for the ORFs ensure that the forecasts remain valid for arbitrary sky distributions of stars and pulsars, partial-sky coverage, and realistic survey configurations.  
In particular, the joint inclusion of astrometric and PTA measurements can substantially enhance sensitivity to both isotropic and kinematic dipolar components of the SGWB, with the degree of improvement strongly depending on the angular distribution of observed sources.

{In the following, we extend the analyses of \cite{Cruz:2024diu,Cruz:2024esk,Cruz:2024svc} by incorporating the most recent specifications for the so-called AA4 configuration of the SKAO, see \cite{braun2019anticipatedperformancesquarekilometre}. We also go beyond the weak signal approximation made in~\cite{Cruz:2024esk,Cruz:2024svc}, based on a Gaussian likelihood in the intensity, which allowed for simple, analytical expressions for the expected measurement precision of the GW properties such as the dipole amplitude. However, since the weak signal approximation will not hold in the SKAO era, we instead follow the approach used in~\cite{Cruz:2024diu}, using a Gaussian likelihood in the timing residuals and astrometric deflections, which is also more in line with standard PTA analyses~\citep{Anholm:2008wy}}

We assume a total of $N_{\rm psr}=200$ monitored millisecond pulsars, a cadence of 14~days, and a total observation time of 10~years.  
The projected sensitivity of the SKAO in terms of GW characteristic strain is expected to be roughly an order of magnitude better than the average achieved by the NANOGrav 15-year dataset.  
Nevertheless, we also explore the potential gains achievable under improved sensitivity scenarios, with the goal of assessing whether enhancements beyond the AA4 configuration could be scientifically warranted and beneficial for probing kinematic anisotropies in the SGWB.

\subsection{Forecasts for pulsar timing measurements}
\label{sec_forpta}


We start with a Gaussian likelihood for the timing residuals $\delta t = r + n$ , where $r$ is GW induced residual and $n$ the noise contribution,
\begin{align}
        -2 \ln L = \sum_f \vec{\delta t}\cdot C^{-1}\cdot  \vec{\delta t}^T + \ln \det C + \frac{N_{\rm \textbf{psr}}}{2}\ln 2\pi\,,
\end{align}
and $N_{\rm \textbf{psr}}$ is the number of pulsars. The noise is taken to be zero mean, Gaussian distributed and uncorrelated across different pulsars $\langle n_p(f) n_q(f') \rangle = \delta_{pq}\delta(f-f')P_n(f)$. 
Thus, the covariance matrix of the timing residuals depends on frequency and  has the form
\begin{align}
    \label{eq:cov_pta}
    C_{ab} = \frac{3 H_0^2}{32\pi^2 f^3}\, \Omega_{\rm GW}(f)\, \Gamma_{ab}(f) + \delta_{ab} P_n(f)\,,
\end{align}
where $\Gamma_{ab}(f)$ is the overlap function in the presence of kinematic anisotropies (see eq.~\eqref{eq:Orf_upto_dipole}), expanded up to the dipole term. In this case the Fisher matrix reduces to 
\begin{align}
F_{\alpha\beta} =  
\mathrm{Tr}\!\left[
\frac{\partial C_{ab}}{\partial \Theta_\alpha}\,
C_{ab}^{-1}\,
\frac{\partial C_{ab}}{\partial \Theta_\beta}\,
C_{ab}^{-1}
\right].
\label{def_fisher}
\end{align}

\begin{table}[h!]
\centering
\begin{tabular}{lccc}
		\hline
PTA configuration  & $\log A_{\rm GW}$ & $\gamma$  &  $\beta$ \\
		\hline
Current PTA (NG15)  & $-14.16 ^{+0.22}_{-0.23}$ &  $3.3 ^{+0.58}_{-0.57}$ & $ < 0.297$ \\
SKAO   (200) & $-14.2 \pm 0.0151 $ & $3.2 \pm {0.08726}$ & $0.00123 \pm {0.07127}$\\
Improved SKAO   (1000) & $-14.2 \pm {0.0098}$ & $3.2 \pm {0.03811}$ & $0.00123 \pm {0.03922}$ \\
		\hline
\end{tabular}
\caption{Comparison between current PTA versus SKAO  sensitivity to the properties of the SGWB at $95\%$ C.L.} 
\label{tab_table1}
\end{table}

\begin{figure}[t!]
    \centering
	\includegraphics[width=0.9 \columnwidth]{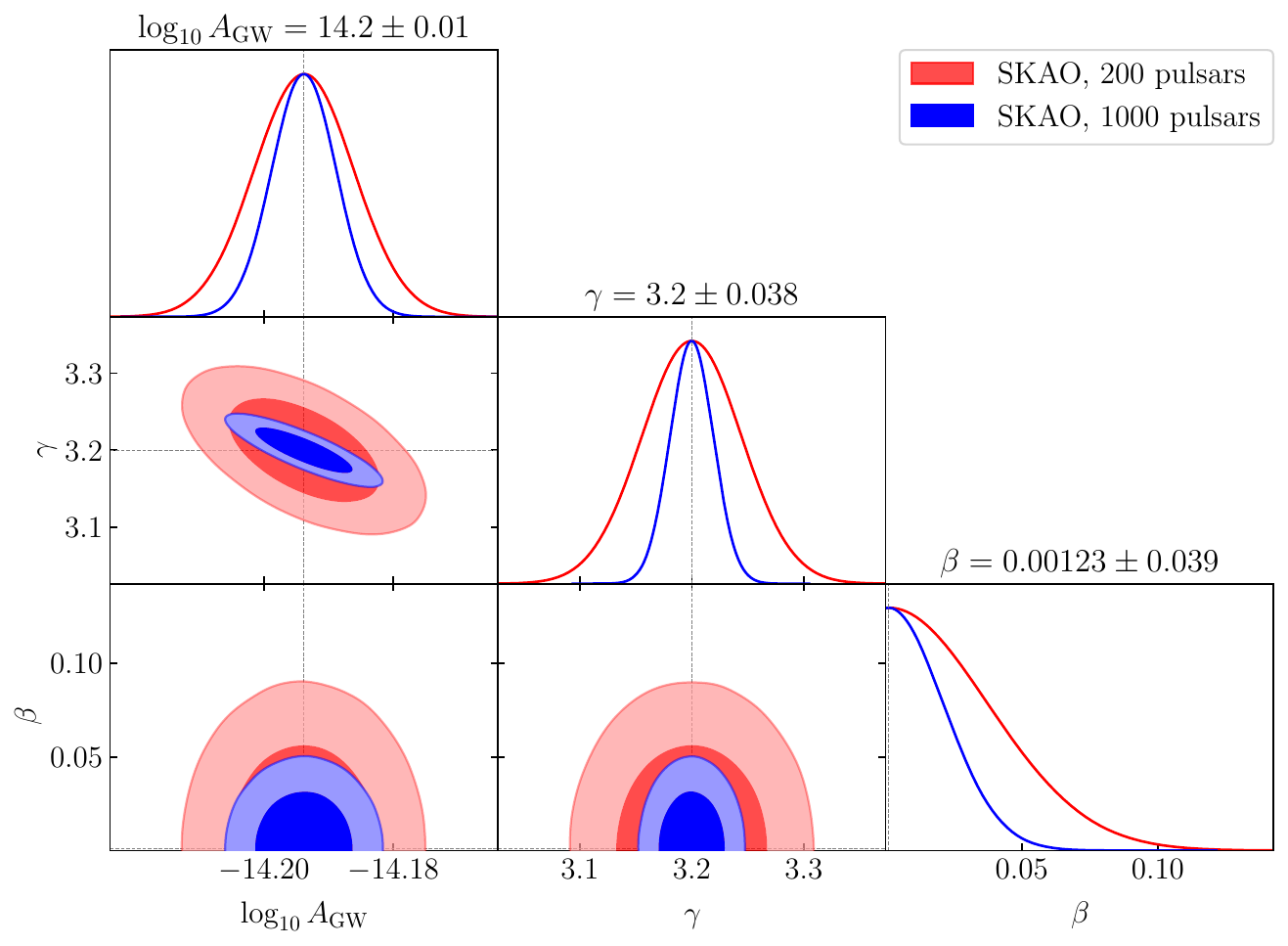}
    \caption{Fisher forecast for $\log A_{\rm GW}$, $\gamma$ and $\beta$ at $95\%$ C. L. for the {two} cases described in section \ref{sec_forpta}. 
    }
    \label{fig_1}
\end{figure}

We applied the Fisher matrix formalism to {two} representative scenarios, with the results summarized in Table~\ref{tab_table1} and Fig.~\ref{fig_1}. In the first case, we considered {the SKAO–AA4 configuration described earlier, which assumes a tenfold improvement in sensitivity relative to the current PTA noise levels (in terms of characteristic strain),} consisting of {200} pulsars with identical noise across different pulsars, adopting a noise curve generated with \texttt{Hasasia}~\citep{Hazboun2019Hasasia}. The second case corresponds to a futuristic scenario with 1000 pulsars, representative of the potential capabilities of SKAO after 20~years of observations, while maintaining the same noise levels as in the SKAO–AA4 case.

Table~\ref{tab_table1} demonstrates that the SKAO will deliver a substantial enhancement in sensitivity compared to existing PTA datasets. In the AA4 configuration, the expected uncertainties on the parameter $\beta$ are reduced by roughly a factor of four, while in the more ambitious futuristic scenario—featuring 1000 monitored pulsars—the improvement reaches about an order of magnitude. These gains represent a major step forward in PTA capabilities. However, they still fall short of the precision required for a statistically significant detection of kinematic anisotropies, which would demand at least an additional order-of-magnitude increase in overall sensitivity. SKAO's much lower noise levels compared to current PTA configurations imply that it will be operating in the strong signal regime, where the variance of the signal itself dominates the covariance (see eq.~\ref{eq:cov_pta}). Thus, further reductions in the noise level are not expected to provide any improvements. In this regime, the forecast error scales as 
\begin{align}
    \Delta \beta \propto \sqrt{\frac{1}{T_{\rm obs}N_{\rm psr}} }\,,
\end{align}
implying that the required gain in sensitivity can only come from large increases in $T_{\rm obs}$ and $N_{\rm psr}$, well beyond the already ambitious $N_{\rm psr}=1000$ case already considered here.

Motivated by this limitation, we next explore how combining PTA observations with independent astrometric measurements could further enhance sensitivity to such effects.

\subsection{Forecasts for synergies between pulsar timing and astrometric
measurements}
\label{sec_forastro}

We now turn to the case of astrometric observations alone.  
The two-point correlation of stellar deflections introduced in Sec.~\ref{sec_astronly} can be expanded in a suitable basis with coefficients $p_n$,  
\begin{align}
\label{est2ddef}
\langle \delta n_a^i \, \delta n_b^j \rangle 
\equiv C^{ij}_{ab} 
= \sum_{n=0,1} p_n\, H_{ab,n}^{ij} + N_{ab}^{ij}.
\end{align}
Here, $a$ and $b$ label the two stars being correlated, while $i$ and $j$ denote the spatial vector components.  
The index $n$ runs over the monopole ($n=0$) and dipole ($n=1$) contributions to the signal.  
Instrumental and measurement uncertainties are described by the noise matrix $N_{ab}^{ij}$.  
The coefficients $p_n$ and kernels $H_{ab,n}^{ij}$ encapsulate the properties of the stochastic gravitational-wave background (SGWB) and the corresponding astrometric overlap reduction functions (ORFs).

In the noise dominated limit with a diagonal noise covariance matrix \mbox{$N_{ab}^{ij} = \sigma^2_N \delta_{ab}\delta^{ij}$}, we obtain the Fisher matrix $F$ for 
the parameters $p_0, p_1$ controlling 
the monopole and dipole correlations:
\renewcommand{\arraystretch}{1.5}
\begin{align}
\label{eq:astro_fisher}
    F = \frac{1}{2 \sigma^4_N}\begin{pmatrix}
       \Tr[\hb_{0}\hb_{0}]\quad  &  \Tr[\hb_{0}\hb_{1}] \\
       \Tr[\hb_{0}\hb_{1}]\quad & \Tr[\hb_{1}\hb_{1}]
    \end{pmatrix}\,.
\end{align}
The statistical information contained in the coefficients $(p_0, p_1)$ can be translated into constraints on the physical parameters $(A_{\rm GW}, \gamma, \beta)$, which describe the amplitude, spectral slope, and kinematic anisotropies of the SGWB.  
A detailed derivation of this mapping is provided in Ref.~\cite{Cruz:2024diu}, which also presents the explicit Fisher-matrix formalism for joint analyses combining astrometric deflection data with pulsar timing array (PTA) measurements.  
That work further discusses practical computational strategies for efficiently handling the large and highly correlated covariance matrices that arise in analyses involving a vast number of stars, ensuring both numerical stability and scalability of the inference pipeline.  
The technical aspects of these calculations lie beyond the scope of the present discussion, and in what follows we focus on the results obtained within that framework.


\begin{figure}[t!]
    \centering
	\includegraphics[width=0.9 \columnwidth]{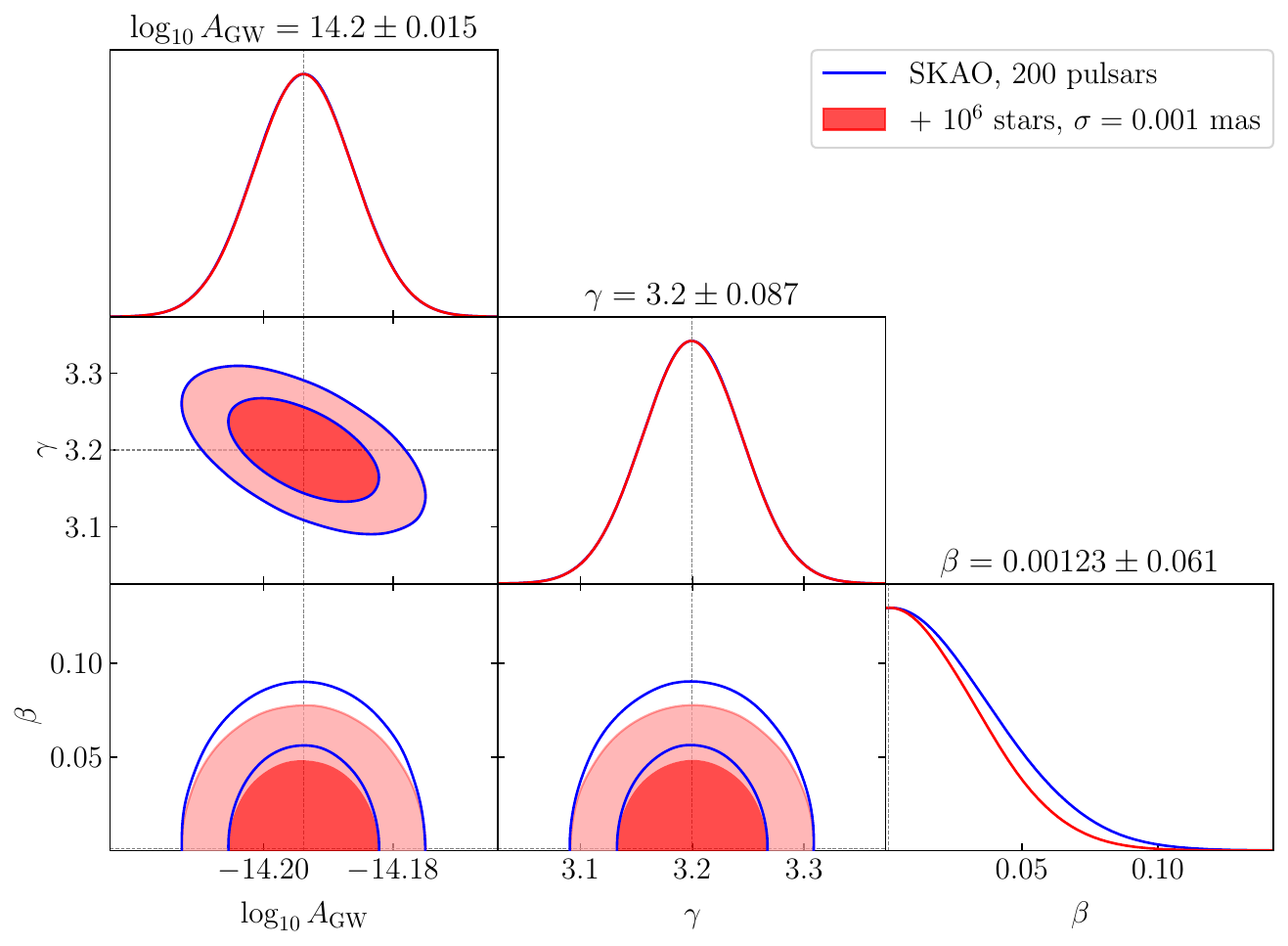}
    \caption{
    Fisher forecast for $\log A_{\rm GW}$, $\gamma$ and $\beta$ at $95\%$ C. L. analysing  pulsar timing residuals correlated
    with astrometry, as described in Section~\ref{sec_forastro}. } 
    \label{fig_2}
\end{figure}

\begin{table}[h!]
\centering
\begin{tabular}{lccc}
		\hline
PTA+astrometry configuration  & $\log A_{\rm GW}$ & $\gamma$  &  $\beta$ \\
		\hline
SKAO (200) & $-14.2 \pm {0.0151}$ & $3.2 \pm {0.08726} $ & $0.00123 \pm {0.07127}$\\
SKAO (200) + Astrometry ($10^{6}$)& $-14.2 \pm {0.0151}$ & $3.2 \pm {0.08726}$ & $0.00123 \pm {0.06119}$ \\
		\hline
\end{tabular}
\caption{Comparison of SKAO forecast sensitivity and SKAO cross correlated with
astrometry at $95\%$ C.L.
}
\label{tab_table2}
\end{table}

Table~\ref{tab_table2} and 
figure~\ref{fig_2} present a comparison between the sensitivity forecasts for the SKAO--AA4 configuration and those obtained from a joint analysis that cross-correlates SKAO data with astrometric measurements of $10^{6}$ stars.  
For the latter, we assume identical noise levels for the astrometric deflection of all stars, given by $\sigma_S^2 = 2 [\Delta\theta_{\rm rms}]^2 T_{\rm cad}$, where we adopt $\Delta\theta_{\rm rms} = 0.001~\mathrm{mas}$,  and $T_{\rm cad} = \mathrm{year}/15$. The chosen noise levels are lower than those expected for \emph{Gaia} DR4 and DR5 (for \emph{Gaia} noise levels we see no noticeable improvement), while the cadence corresponds approximately to its observational cadence.\footnote{\url{https://www.cosmos.esa.int/web/gaia/science-performance}.}
The results demonstrate that including astrometric correlations leads to a measurable enhancement in constraining power for $\beta$, with an improvement of roughly $\mathcal{O}(10\%)$ over the SKAO-only case. {On the other hand, there is no visible improvement for the isotropic signal parameters, likely due to the high precision SKAO already provides for these, thus the relative improvements brought about by including astrometry are tiny compared to SKAO alone. The additional sky coverage provided by the much larger number of stars might also help in detection of anisotropy, leading to better constraints on $\beta$ from the joint analyses.}

Although modest, this gain is significant given the already improved sensitivity of the SKAO configuration with respect to current PTA measurements,  and highlights the potential of combining PTA and future astrometric data to further refine measurements of the SGWB.
Nevertheless, even such combined analyses are not yet sufficient to achieve the sensitivity required for a statistically significant detection of the expected kinematic anisotropies in the SGWB, indicating that genuinely new observational strategies or analysis techniques will be needed to reach this goal.

\section{Conclusions}
\label{sec_conclusions}

In this chapter, we have reviewed and extended our recent works
\cite{Tasinato:2023zcg,Cruz:2024svc,Cruz:2024esk,Cruz:2024diu}
on probing kinematic anisotropies in the stochastic gravitational-wave background (SGWB) with pulsar timing arrays (PTAs).  
We have applied these techniques to realistic forecasts for the Square Kilometre Array Observatory (SKAO), considering both the standard AA4 configuration monitoring $\sim 200$ pulsars and a more optimistic scenario with $\sim 1000$ pulsars.

Our analysis shows that SKAO will significantly improve constraints on kinematic anisotropies relative to existing PTA datasets.  
However, even in the most optimistic SKAO configuration, the expected sensitivity remains insufficient to detect the level of anisotropy predicted from kinematic effects, which are expected to be suppressed by $\beta \sim 10^{-3}$ relative to the isotropic SGWB amplitude.  

\begin{figure}[t!]
    \centering
	\includegraphics[width=0.6 \columnwidth]{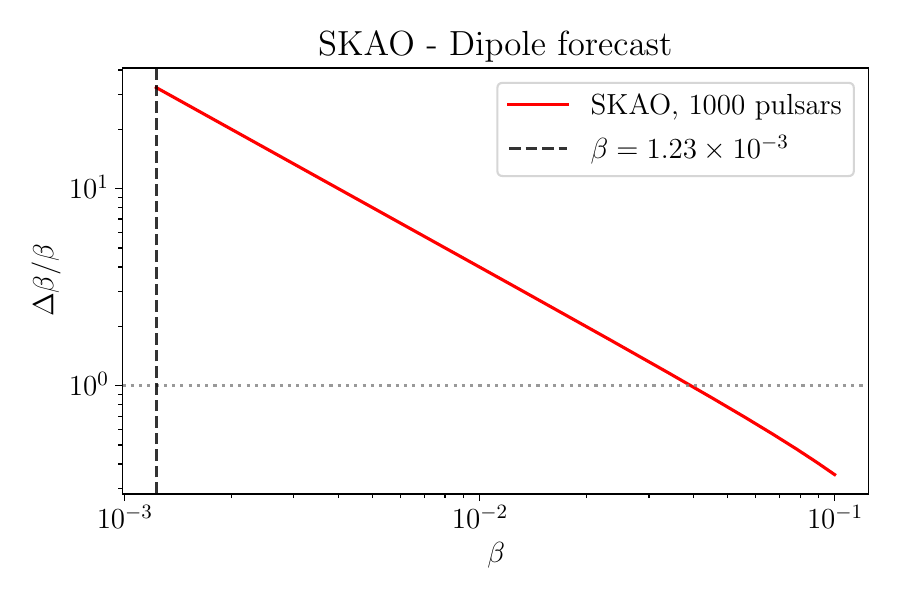}
    \caption{Fisher forecast for the fractional error (at $95\%$ C.L) on the dipole for enhanced SKAO with 1000 pulsars and 20 years of observation. The fiducial value of $\beta$ is varied between the current value measured with the CMB $1.23 \times 10^{-3}$ (black dashed line) and $0.1$. The area below the grey dotted line corresponds to the minimum level of kinematic dipole that will be detectable with SKAO setup, where $\Delta \beta/ \beta <1$.}
    \label{fig_3}
\end{figure}

Motivated by this limitation, we explored the potential of augmenting PTA data with astrometric measurements, using \emph{Gaia}-like specifications as a benchmark.  
Cross-correlating SKAO timing data with astrometric deflections from $10^{6}$ stars yields an additional improvement in sensitivity, but still does not reach the level required for a statistically significant detection of kinematic anisotropies.  
This suggests that further methodological development --- or the incorporation of additional observables --- will be necessary to fully unlock this science case.  
For example, future cross-correlation strategies that include complementary probes, such as galaxy shimmering~\citep{Mentasti:2024fgt}, may offer promising avenues to enhance sensitivity.


To conclude on a positive and forward-looking note, we quantified the minimum anisotropy amplitude detectable with SKAO configurations.  
As shown in Fig.~\ref{fig_3}, an anisotropic SGWB contribution with fiducial amplitude $\beta \gtrsim 4\times 10^{-2}$ would be confidently detected by the enhanced SKAO scenario considered here. 
{Thus, if early-Universe mechanisms imprint directional structure in the GW background—through tensor-sector symmetry breaking or vector-driven anisotropy \citep{Maleknejad:2012fw}, possibly motivated  by CMB directional anomalies \citep{Aluri:2022hzs}—SKAO could detect these signatures with high significance.
More in general, this also reflects the fact that SKAO's ability to detect anisotropies, kinematic or otherwise, will be around the $10^{-2}$ level.}
The search for  {SGWB} anisotropies therefore remains a compelling target for future PTA science, motivating continued theoretical and data-analysis innovation.

\subsection*{Acknowledgments}
 We are partially funded by the STFC
grants ST/T000813/1 and ST/X000648/1.











\bibliographystyle{abbrvnat-maxbibnames4}
\bibliography{GWBKAchapter} 

\end{document}

%% file: journal-names.tex
\newcommand{\actaa}{Acta Astron.} 
\newcommand{\araa}{ARA\&A} 
\newcommand{\aar}{A\&ARv} 
\newcommand{\aapr}{A\&ARv} 
\newcommand{\ab}{Astrobiol.} 
\newcommand{\aj}{AJ} 
\newcommand{\apj}{ApJ} 
\newcommand{\apjl}{ApJL} 
\newcommand{\apjs}{ApJSS} 
\newcommand{\ao}{Appl. Opt.} 
\newcommand{\apss}{Astro. \& Space Sci.} 
\newcommand{\aap}{A\&A} 
\newcommand{\aaps}{A\&AS.} 
\newcommand{\baas}{Bull. Am. Astron. Soc.} 
\newcommand{\caa}{Chinese A\&A} 
\newcommand{\cjaa}{Chinese J. A\&A} 
\newcommand{\cqg}{Class. Quantum Gravity} 
\newcommand{\gal}{Galaxies} 
\newcommand{\gca}{Geo. Cosmo. Acta} 
\newcommand{\icarus}{Icarus} 
\newcommand{\jcap}{JCAP} 
\newcommand{\jgr}{J. Geophys. Res.} 
\newcommand{\jgrp}{J. Geophys. Res. Planets} 
\newcommand{\jqsrt}{J. Quant. Spectrosc. Radiat. Transf.} 
\newcommand{\memsai}{Mem. SAIt} 
\newcommand{\mnras}{MNRAS} 
\newcommand{\nat}{Nature} 
\newcommand{\nastro}{Nat. Astron.} 
\newcommand{\ncomms}{Nat. Commun.} 
\newcommand{\nphys}{Nat. Phys.} 
\newcommand{\na}{New Astron.} 
\newcommand{\nar}{New Astron. Rev.} 
\newcommand{\physrep}{Phys. Rep.} 
\newcommand{\pra}{Phys. Rev. A} 
\newcommand{\prb}{Phys. Rev. B} 
\newcommand{\prc}{Phys. Rev. C} 
\newcommand{\prd}{Phys. Rev. D} 
\newcommand{\pre}{Phys. Rev. E} 
\newcommand{\prx}{Phys. Rev. X} 
\newcommand{\prl}{Phys. Rev. Let.} 
\newcommand{\psj}{Planet. Sci. J.} 
\newcommand{\planss}{Planet. Space Sci.} 
\newcommand{\pnas}{Proc. Natl Acad. Sci. USA} 
\newcommand{\procspie}{Proc. SPIE} 
\newcommand{\pasa}{PASA} 
\newcommand{\pasj}{PASJ} 
\newcommand{\pasp}{PASP} 
\newcommand{\rmxaa}{RMXAA} 
\newcommand{\sci}{Science} 
\newcommand{\sciadv}{Sci. Adv.} 
\newcommand{\solphys}{Sol. Phys.} 
\newcommand{\sovast}{Soviet Ast.} 
\newcommand{\ssr}{Space Sci. Rev.} 
\newcommand{\uni}{Universe} 